\documentclass{aa}  

\usepackage{graphicx}
\usepackage{txfonts}
\usepackage{color}
\usepackage{multirow}

\newcommand{\tablenotea}[1]{\parbox{16.0cm}{\indent \footnotesize{#1}}}
\newcommand{\tablenoteb}[1]{\parbox{7.1cm}{\indent \footnotesize{#1}}}

\newcommand{\ieeetap}{IEEE Trans. Antennas Propag.}

\newcommand{\jms}{J. Mol. Spectr.}
\newcommand{\jmst}{J. Mol. Struct.}

\newcommand{\jpca}{J. Phys. Chem. A}

\newcommand{\science}{Science}
\newcommand{\sciadv}{Sci. Adv.}
\newcommand{\zpc}{Z. Phys. Chem.}

\begin{document}

\title{Detection of the propargyl radical at $\lambda$\,3 mm\thanks{Based on observations carried out with the Yebes 40m telescope (projects 19A003, 20A014, 20D023, and 21A011) and the IRAM 30m telescope. The 40m radiotelescope at Yebes Observatory is operated by the Spanish Geographic Institute (IGN, Ministerio de Transportes, Movilidad y Agenda Urbana). IRAM is supported by INSU/CNRS (France), MPG (Germany) and IGN (Spain).}}

\titlerunning{Detection of CH$_2$CCH at $\lambda$\,3 mm}
\authorrunning{Ag\'undez et al.}

\author{M.~Ag\'undez\inst{1}, N.~Marcelino\inst{2,3}, C.~Cabezas\inst{1}, R.~Fuentetaja\inst{1}, B.~Tercero\inst{2,3}, P.~de~Vicente\inst{3}, \and J.~Cernicharo\inst{1}}

\institute{
Instituto de F\'isica Fundamental, CSIC, Calle Serrano 123, E-28006 Madrid, Spain\\ \email{marcelino.agundez@csic.es, jose.cernicharo@csic.es} \and
Observatorio Astron\'omico Nacional, IGN, Calle Alfonso XII 3, E-28014 Madrid, Spain \and
Observatorio de Yebes, IGN, Cerro de la Palera s/n, E-19141 Yebes, Guadalajara, Spain
}

\date{Received; accepted}

 
\abstract
{We report the detection of the propargyl radical (CH$_2$CCH) in the cold dark cloud \mbox{TMC-1} in the $\lambda$\,3 mm wavelength band. We recently discovered this species in space toward the same source at a wavelength of $\lambda$\,8 mm. In those observations, various hyperfine components of the 2$_{0,2}$-1$_{0,1}$ rotational transition, at 37.5 GHz, were detected using the Yebes 40m telescope. Here, we used the IRAM 30m telescope to detect ten hyperfine components of the 5$_{0,5}$-4$_{0,4}$ rotational transition, lying at 93.6 GHz. The observed frequencies differ by 0.2 MHz with respect to the predictions from available laboratory data. This difference is significant for a radioastronomical search for CH$_2$CCH in interstellar sources with narrow lines. We thus included the measured frequencies in a new spectroscopic analysis to provide accurate frequency predictions for the interstellar search for propargyl at mm wavelengths. Moreover, we recommend that future searches for CH$_2$CCH in cold interstellar clouds are carried out at $\lambda$\,3 mm, rather than at $\lambda$\,8 mm. The 5$_{0,5}$-4$_{0,4}$ transition is about five times more intense than the 2$_{0,2}$-1$_{0,1}$ one in \mbox{TMC-1}, which implies that detecting the former requires about seven times less telescope time than detecting the latter. We constrain the rotational temperature of CH$_2$CCH in \mbox{TMC-1} to 9.9 $\pm$ 1.5 K, which indicates that the rotational levels of this species are thermalized at the gas kinetic temperature. The revised value of the column density of CH$_2$CCH (including ortho and para species) is (1.0\,$\pm$\,0.2)\,$\times$\,10$^{14}$ cm$^{-2}$, and thus the CH$_2$CCH/CH$_3$CCH abundance ratio is revised from slightly below one to nearly one. This study opens the door for future detections of CH$_2$CCH in other cold interstellar clouds, making possible to further investigate the role of this very abundant hydrocarbon radical in the synthesis of large organic molecules such as aromatic rings.}


\maketitle

\section{Introduction}

The on going line surveys of the Taurus Molecular Cloud 1 (\mbox{TMC-1}) at the Green Bank telescope (GOTHAM; \citealt{McGuire2020}) and at the Yebes 40m telescope (QUIJOTE; \citealt{Cernicharo2021d}), are demonstrating that complex hydrocarbons, including cyclic and polycyclic aromatic ones, are formed in situ in cold dense clouds. Examples of such molecules detected toward \mbox{TMC-1} are propylene (CH$_2$CHCH$_3$), vinyl and allenyl acetylene (CH$_2$CHCCH and CH$_2$CCHCCH), cyclopentadiene ($c$-C$_5$H$_6$), indene ($c$-C$_9$H$_8$), and benzyne ($o$-C$_6$H$_4$) \citep{Marcelino2007,Cernicharo2021a,Cernicharo2021b,Cernicharo2021c,Cernicharo2021d,Burkhardt2021}. Moreover, ethyl acetylene (CH$_3$CH$_2$CCH) and ethynyl benzene ($c$-C$_6$H$_5$CCH) have been tentatively detected \citep{Cernicharo2021a,Cernicharo2021e}, and there is strong evidence for the presence of aromatic rings such as benzene and naphthalene from the detection of their CN derivatives \citep{McGuire2018,McGuire2021,Cernicharo2021e}.

It is not yet well understood which chemical routes are behind the formation of these aromatic cycles in cold dark clouds like \mbox{TMC-1}. Hydrocarbon radicals are likely key players in the synthesis of these large molecules from smaller species. However, only a few such radicals have been detected. The methylidyne radical CH and the polyacetylenic radicals C$_2$H, C$_3$H, C$_4$H, and even longer ones are known from long time ago. Other radicals such as C$_2$H$_3$, C$_3$H$_3$, or C$_3$H$_5$ are likely important pieces in the synthesis of large hydrocarbons but detecting them has been proven to be difficult due to different possible reasons like spectral dilution due to splitting of rotational lines into numerous fine and hyperfine components, low abundance, low dipole moment, or lack of sufficiently sensitive radioastronomical observations. We recently identified the propargyl radical (CH$_2$CCH) toward \mbox{TMC-1} as part of the QUIJOTE line survey \citep{Agundez2021a}. It was found that CH$_2$CCH is one of the most abundant radicals in \mbox{TMC-1}, being present at the level of its closed-shell counterpart CH$_3$CCH. Being that abundant, the propargyl radical becomes a very attractive candidate to play an important role in the synthesis of aromatic molecules. For example, calculations indicate that the propargyl radical self reaction can lead to cyclization producing the aromatic radical phenyl radical at low temperatures \citep{Miller2001,Zhao2021}.

\begin{figure*}
\centering
\includegraphics[angle=0,width=\textwidth]{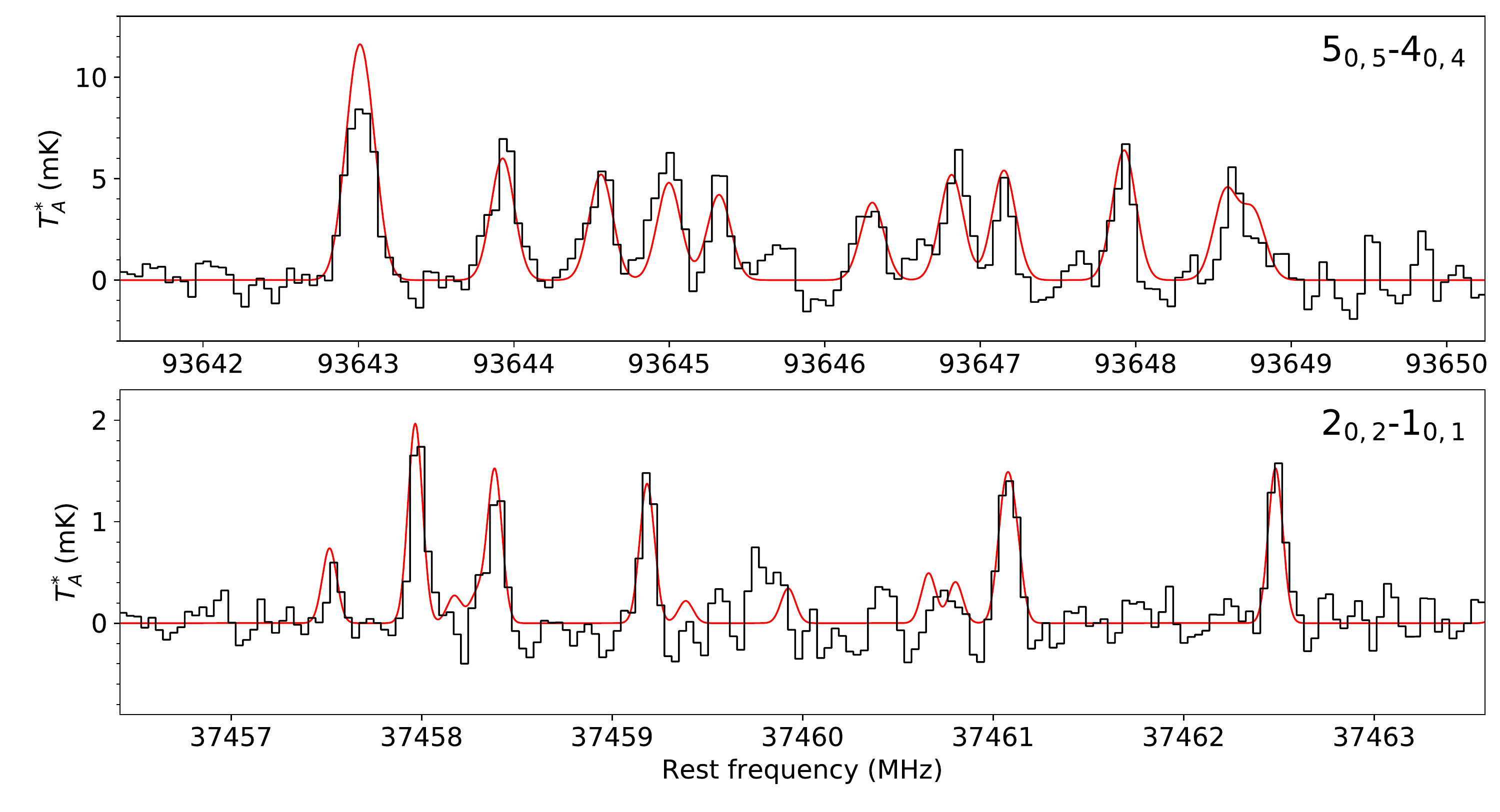}
\caption{Observed spectra of \mbox{TMC-1} around the 2$_{0,2}$-1$_{0,1}$ and 5$_{0,5}$-4$_{0,4}$ rotational transitions of ortho CH$_2$CCH. The spectrum at 37.5 GHz is taken with the Yebes 40m telescope (black histogram in bottom panel) and that at 93.6 GHz is taken with the IRAM 30m telescope (black histogram in top panel). Transition quantum numbers, frequencies, and derived line parameters are given in Table~\ref{table:lines}. The synthetic spectra (red lines) were computed for a column density of ortho CH$_2$CCH of $7.5\times10^{13}$ cm$^{-2}$, a rotational temperature of 9.9 K, an emission size of 40$''$ of radius, and a linewidth of 0.72 km s$^{-1}$ for the 2$_{0,2}$-1$_{0,1}$ lines and of 0.57 km s$^{-1}$ for the 5$_{0,5}$-4$_{0,4}$ lines (see text).} \label{fig:lines}
\end{figure*}

The radical CH$_2$CCH was detected in \mbox{TMC-1} at $\lambda$\,8 mm through six hyperfine components belonging to the 2$_{0,2}$-1$_{0,1}$ rotational transition. Here we report the detection of CH$_2$CCH toward \mbox{TMC-1} at $\lambda$\,3 mm. We observed the 5$_{0,5}$-4$_{0,4}$ transition in ten hyperfine components, with frequencies that differ by 0.2 MHz from previous available predictions. We thus used the observed frequencies to improve the spectroscopic parameters of CH$_2$CCH and provide accurate predictions to guide future astronomical searches. Moreover, the $\lambda$\,3 mm line is about five times more intense than the $\lambda$\,8 mm one, which suggests that the search for CH$_2$CCH in other cold dark clouds is more favorable in the $\lambda$\,3 mm wavelength band.

\begin{table*}
\small
\caption{Observed line parameters of CH$_2$CCH in \mbox{TMC-1}.}
\label{table:lines}
\centering
\begin{tabular}{lrrcccr}
\hline \hline
\multicolumn{1}{c}{Transition\,$^a$} & \multicolumn{1}{c}{$\nu_{\rm calc}$\,$^b$} & \multicolumn{1}{c}{$\nu_{\rm obs}$\,$^c$} & \multicolumn{1}{c}{$\Delta v$\,$^d$}      & \multicolumn{1}{c}{$T_A^*$ peak} & \multicolumn{1}{c}{$\int T_A^* dv$} & \multicolumn{1}{c}{S/N\,$^e$} \\
\multicolumn{1}{c}{} & \multicolumn{1}{c}{(MHz)}        & \multicolumn{1}{c}{(MHz)}        & \multicolumn{1}{c}{(km s$^{-1}$)}  & \multicolumn{1}{c}{(mK)}                   & \multicolumn{1}{c}{(mK km s$^{-1}$)} & \multicolumn{1}{c}{($\sigma$)} \\
\hline
2$_{0,2}$-1$_{0,1}$ \hspace{0.1cm} $J$ = 3/2-1/2 \hspace{0.26cm} $F_1$ = 2-1 \hspace{0.1cm} $F$ = 2-1 & 37457.515 & 37457.542(20) & 0.55(35) & 0.60 & 0.35(22) & 4.5 \\
2$_{0,2}$-1$_{0,1}$ \hspace{0.1cm} $J$ = 5/2-3/2 \hspace{0.26cm} $F_1$ = 3-2 \hspace{0.1cm} $F$ = 4-3 & 37457.965 & 37457.981(10) & 0.71(16) & 1.92 & 1.45(27) & 16.4 \\
2$_{0,2}$-1$_{0,1}$ \hspace{0.1cm} $J$ = 5/2-3/2 \hspace{0.26cm} $F_1$ = 2-1 \hspace{0.1cm} $F$ = 3-2 & 37458.382 & 37458.390(13) & 0.80(27) & 1.28 & 1.09(29) & 11.6 \\
2$_{0,2}$-1$_{0,1}$ \hspace{0.1cm} $J$ = 5/2-3/2 \hspace{0.26cm} $F_1$ = 3-2 \hspace{0.1cm} $F$ = 3-2 & 37459.186 & 37459.187(10) & 0.65(14) & 1.58 & 1.09(22)\,$^f$ & 12.9 \\
2$_{0,2}$-1$_{0,1}$ \hspace{0.1cm} $J$ = 5/2-3/2 \hspace{0.26cm} $F_1$ = 2-1 \hspace{0.1cm} $F$ = 2-1 & 37461.057 & \multirow{2}{*}{\bigg\}37461.078(11)} & \multirow{2}{*}{0.85(15)} & \multirow{2}{*}{1.53} & \multirow{2}{*}{1.38(25)\,$^g$} & \multirow{2}{*}{14.3} \\
2$_{0,2}$-1$_{0,1}$ \hspace{0.1cm} $J$ = 5/2-3/2 \hspace{0.26cm} $F_1$ = 2-1 \hspace{0.1cm} $F$ = 2-1 & 37461.111 & & & & & \\
2$_{0,2}$-1$_{0,1}$ \hspace{0.1cm} $J$ = 3/2-1/2 \hspace{0.26cm} $F_1$ = 2-1 \hspace{0.1cm} $F$ = 3-2 & 37462.481 & 37462.489(10) & 0.76(17) & 1.63 & 1.31(25) & 14.3 \\
& & & & & & \\
5$_{0,5}$-4$_{0,4}$ \hspace{0.1cm} $J$ = 11/2-9/2 \hspace{0.1cm} $F_1$ = 6-5 \hspace{0.1cm} $F$ = 7-6 & 93642.970 & \multirow{2}{*}{\bigg\}93643.011(15)} & \multirow{2}{*}{0.69(11)} & \multirow{2}{*}{9.00} & \multirow{2}{*}{6.60(93)\,$^g$} & \multirow{2}{*}{22.3} \\
5$_{0,5}$-4$_{0,4}$ \hspace{0.1cm} $J$ = 11/2-9/2 \hspace{0.1cm} $F_1$ = 5-4 \hspace{0.1cm} $F$ = 6-5 & 93643.059 & & & & & \\
5$_{0,5}$-4$_{0,4}$ \hspace{0.1cm} $J$ = 11/2-9/2 \hspace{0.1cm} $F_1$ = 6-5 \hspace{0.1cm} $F$ = 6-5 & 93643.927 & 93643.940(22) & 0.65(18) & 6.25 & 4.32(97) & 15.1 \\
5$_{0,5}$-4$_{0,4}$ \hspace{0.1cm} $J$ = 11/2-9/2 \hspace{0.1cm} $F_1$ = 6-5 \hspace{0.1cm} $F$ = 5-4 & 93644.560 & 93644.557(28) & 0.66(24) & 4.98 & 3.51(99) & 12.1 \\
5$_{0,5}$-4$_{0,4}$ \hspace{0.1cm} $J$ = 11/2-9/2 \hspace{0.1cm} $F_1$ = 5-4 \hspace{0.1cm} $F$ = 5-4 & 93644.997 & 93644.982(22) & 0.65(16) & 6.19 & 4.30(92) & 15.0 \\
5$_{0,5}$-4$_{0,4}$ \hspace{0.1cm} $J$ = 11/2-9/2 \hspace{0.1cm} $F_1$ = 5-4 \hspace{0.1cm} $F$ = 4-3 & 93645.319 & 93645.325(18) & 0.38(14) & 5.76 & 2.34(71) & 10.7 \\
5$_{0,5}$-4$_{0,4}$ \hspace{0.1cm} $J$ = 9/2-7/2 \hspace{0.26cm}  $F_1$ = 4-3 \hspace{0.1cm} $F$ = 4-3 & 93646.305 & 93646.286(48) & 0.61(30) & 3.68 & 2.40(114) & 8.6 \\
5$_{0,5}$-4$_{0,4}$ \hspace{0.1cm} $J$ = 9/2-7/2 \hspace{0.26cm}  $F_1$ = 4-3 \hspace{0.1cm} $F$ = 5-4 & 93646.815 & 93646.849(31) & 0.61(32) & 5.73 & 3.72(136) & 13.4 \\
5$_{0,5}$-4$_{0,4}$ \hspace{0.1cm} $J$ = 9/2-7/2 \hspace{0.26cm}  $F_1$ = 5-4 \hspace{0.1cm} $F$ = 5-4 & 93647.152 & 93647.155(25) & 0.35(17) & 5.13 & 1.89(87) & 9.0 \\
5$_{0,5}$-4$_{0,4}$ \hspace{0.1cm} $J$ = 9/2-7/2 \hspace{0.26cm}  $F_1$ = 5-4 \hspace{0.1cm} $F$ = 6-5 & 93647.925 & 93647.919(24) & 0.44(18) & 6.36 & 2.99(103) & 12.7 \\
5$_{0,5}$-4$_{0,4}$ \hspace{0.1cm} $J$ = 9/2-7/2 \hspace{0.26cm}  $F_1$ = 5-4 \hspace{0.1cm} $F$ = 4-3 & 93648.580 & \multirow{2}{*}{\bigg\}93648.648(41)} & \multirow{2}{*}{0.62(45)} & \multirow{2}{*}{4.65} & \multirow{2}{*}{3.07(145)\,$^g$} & \multirow{2}{*}{11.0} \\
5$_{0,5}$-4$_{0,4}$ \hspace{0.1cm} $J$ = 9/2-7/2 \hspace{0.26cm}  $F_1$ = 4-3 \hspace{0.1cm} $F$ = 3-2 & 93648.750 & & & & & \\
\hline
\end{tabular}
\tablenotea{\\
The line parameters $\nu_{\rm obs}$, $\Delta v$, $T_A^*$ peak, and $\int T_A^* dv$ as well as the associated errors were derived from a Gaussian fit to each line profile.\\
$^a$\,Quantum numbers from the coupling scheme of \cite{Tanaka1997}.\\
$^b$\,Calculated frequencies $\nu_{\rm calc}$ from the combined laboratory+astronomical fit carried out in this work.\\
$^c$\,Observed frequencies adopting a systemic velocity of 5.83 km s$^{-1}$ for \mbox{TMC-1} \citep{Cernicharo2020}.\\
$^d$\,$\Delta v$ is the full width at half maximum (FWHM).\\
$^e$\,The signal-to-noise ratio is computed as S/N = $\int T_A^* dv$ / [rms $\times$ $\sqrt{\Delta v \times \delta \nu (c / \nu_{\rm calc})}$], where $c$ is the speed of light and $\delta \nu$ is the spectral resolution. For the lines observed with the Yebes 40m telescope at 37.5 GHz $\delta \nu$\,=\,0.03815 MHz and rms\,=\,0.19 mK, while for the lines observed with the IRAM 30m telescope at 93.6 GHz $\delta \nu$\,=\,0.04884 MHz and rms\,=\,0.9 mK. The rest of parameters are given in the table.\\
$^f$\,Line overlaps with the 2$_{1,2}$-1$_{1,1}$ transition of $syn$-C$_2$H$_3$OH, which lies at 37459.184 MHz (see \citealt{Agundez2021b}). The observed intensity is thus the sum of the CH$_2$CCH and the $syn$-C$_2$H$_3$OH lines.\\
$^g$\,Observed line results from a blend of two unresolved hyperfine components.
}
\end{table*}

\section{Observations}

The observations of \mbox{TMC-1} at $\lambda$\,3 mm were carried out using the IRAM 30m telescope in September 2021. The observed position corresponds to the cyanopolyyne peak of \mbox{TMC-1}, $\alpha_{J2000}=4^{\rm h} 41^{\rm  m} 41.9^{\rm s}$ and $\delta_{J2000}=+25^\circ 41' 27.0''$. The 3\,mm EMIR receiver was used connected to a fast Fourier transform spectrometer, providing a spectral resolution of 48.84 kHz. We covered the spectral region around 93.6 GHz, where the 5$_{0,5}$-4$_{0,4}$ rotational transition of CH$_2$CCH is located. We observed two setups at slightly different central frequencies in order to check for spurious signals, line emission from the image band, and other technical artifacts. The observations were performed in the frequency-switching observing mode with a frequency throw of 18\,MHz, large enough to avoid possible contamination from negative frequency-switching artifacts arising from the different hyperfine components of CH$_2$CCH. Pointing scans were performed on strong and nearby quasars every 1-1.5 h, with pointing errors always within 3-5$''$. The antenna focus was checked every $\sim$6 h at the beginning of each observing session and after sunrise. Weather conditions were between good and average for the summer period, with opacities of 0.4-0.5 at 225\,GHz and amounts of precipitable water vapor ranging from 1-3\,mm to 6-7\,mm. The spectra were calibrated in antenna temperature, $T_A^*$, corrected for atmospheric attenuation and for antenna ohmic and spillover losses, using the ATM package \citep{Cernicharo1985,Pardo2001}. The uncertainty in the calibration is estimated to be 10~\%. System temperatures varied between 100 and 140 K and the final $T_A^*$ rms at 93.6 GHz is 1.1 mK after 31.4 h of total on-source telescope time.

The final spectra shown in Fig.~\ref{fig:lines} is obtained after averaging the data taken in September 2021 with previous spectra from our \mbox{TMC-1} 3\,mm line survey \citep{Marcelino2007,Cernicharo2012}. At the frequency of the 5$_{0,5}$-4$_{0,4}$ transition of CH$_2$CCH, the observed time in the survey data is 4.0 h. Including these data has improved the final sensitivity down to 0.9\,mK, resulting in a total on-source integration time of 35.4 h for each polarization (twice this value after averaging the two polarizations).

We also present a more sensitive spectrum of \mbox{TMC-1} at the frequency of the 2$_{0,2}$-1$_{0,1}$ transition of CH$_2$CCH, 37.5 GHz, with respect to that presented by \cite{Agundez2021a}. New data was gathered in several observing sessions between January and May 2021. These data are part of the on going QUIJOTE line survey that is being carried out with the Yebes 40m telescope. The line survey uses a 7 mm receiver covering the Q band, from 31.0 GHz to 50.3 GHz, with horizontal and vertical polarizations. A detailed description of the system is given by \citet{Tercero2021}. Receiver temperatures in the observing sessions carried out during 2020 vary from 22 K at 32 GHz to 42 K at 50 GHz. Some power adaptation in the down-conversion chains has reduced the receiver temperatures during 2021 to 16\,K at 32 GHz and 25\,K at 50 GHz. The backends are $2\times8\times2.5$ GHz fast Fourier transform spectrometers with a spectral resolution of 38.15 kHz providing the whole coverage of the Q band in both polarizations. The QUIJOTE observations are performed using the frequency-switching observing mode with a frequency throw of 10 MHz in the very first observing runs, during November 2019 and February 2020, and of 8 MHz in the later ones. The main beam efficiency of the Yebes 40m telescope varies from 0.6 at 32 GHz to 0.43 at 50 GHz. The intensity scale used in this work, antenna temperature ($T_A^*$), was calibrated using two absorbers at different temperatures and the atmospheric transmission model ATM \citep{Cernicharo1985,Pardo2001}. Calibration uncertainties were adopted to be 10~\%. After including all data taken between November 2019 and May 2021, the total on-source telescope time is 238 h in each polarization (twice this value after averaging the two polarizations). The IRAM 30m and Yebes 40m data were analyzed using the GILDAS software\footnote{\texttt{http://www.iram.fr/IRAMFR/GILDAS/}}.

\section{Results and discussion}

\subsection{Improved rotational spectroscopy for CH$_2$CCH}

The rotational spectrum of the propargyl radical has been measured in the laboratory at frequencies below 38 GHz by \cite{Tanaka1997}. Due to the existence of two equivalent H nuclei, the radical has ortho/para statistics. Ortho levels have $K_a$ even and para levels have $K_a$ odd. The statistical ortho-to-para ratio is three. The dipole moment of CH$_2$CCH has been calculated by \cite{Botschwina1995} to be 0.14~D, while more recently, \cite{Kupper2002} measured a value of 0.150\,$\pm$0.005~D, which is the value we adopt hereafter.

Our IRAM 30m data of \mbox{TMC-1} show a group of lines spanning 6 MHz around 93646 MHz (see top panel in Fig.~\ref{fig:lines}), which we assign to the hyperfine components of the 5$_{0,5}$-4$_{0,4}$ transition of CH$_2$CCH. The measured frequencies are systematically shifted up by 0.2 MHz with respect to the predicted frequencies in the Cologne Database for Molecular Spectroscopy (CDMS; \citealt{Muller2005})\,\footnote{\texttt{https://cdms.astro.uni-koeln.de/}}. The entry in the CDMS is based on a fit to the laboratory frequencies measured by \cite{Tanaka1997}. These authors measured the fine and hyperfine structure of the rotational transitions 1$_{0,1}$-0$_{0,0}$, 2$_{0,2}$-1$_{0,1}$, 2$_{1,2}$-1$_{1,1}$, and 2$_{1,1}$-1$_{1,0}$, lying at 18.7~GHz and in the 37-38~GHz range. Although the experimental accuracy is quite good, a few kilohertz, the limited range of $J$ values covered makes that when extrapolating to the $\lambda$\,3 mm wavelength band, the frequency errors could be significant for radioastronomical purposes. The CDMS quotes frequency errors of just $\sim$\,55 kHz for the hyperfine components of the 5$_{0,5}$-4$_{0,4}$ transition, although our \mbox{TMC-1} observations shows that the error is in fact as high as $\sim$\,200 kHz. This is significant for a radioastronomical search for CH$_2$CCH in sources with narrow lines, such as \mbox{TMC-1}.

In order to obtain more accurate frequency predictions for CH$_2$CCH we carried out a new spectroscopic analysis using the SPFIT program \citep{Pickett1991} including the laboratory frequencies of \cite{Tanaka1997} and the astronomical frequencies measured in \mbox{TMC-1} for the ten hyperfine components of the 5$_{0,5}$-4$_{0,4}$ transition (see derived line parameters in Table~\ref{table:lines}). The Hamiltonian used for the analysis is the same than that employed by \cite{Tanaka1997} and it has the following form:
\begin{equation}
H = H^{rot} + H^{cd} + H^{sr} + H^{mhf}
\end{equation}
where $H^{rot}$ and $H^{cd}$ contain the rotational and centrifugal distortion parameters, respectively, $H^{sr}$ is the spin-rotation term, and $H^{mhf}$ represents the magnetic hyperfine coupling interaction between the unpaired electron and the hydrogen nuclei. A complete description of these terms can be found in \cite{Tanaka1997}. The coupling scheme used is \textbf{J}\,=\,\textbf{N}\,+\,\textbf{S}, \textbf{F}$_1$\,=\,\textbf{J}\,+\,\textbf{I}$_1$, and \textbf{F}\,=\,\textbf{F}$_1$\,+\,\textbf{I}$_2$, where \textbf{I}$_1$\,=\,\textbf{I}$\rm(H_a)$ and \textbf{I}$_2$\,=\,\textbf{I}$\rm(H_{m1})$\,+\,\textbf{I}$\rm(H_{m2})$. The radical CH$_2$CCH has two equivalent H nuclei, the methylenic ones, and the hyperfine interaction term $H^{mhf}$ is thus written explicitly as a two spin system:
\begin{equation}
H^{mhf} = a_F^{\rm(H_a)} \cdot \textbf{S} \cdot \textbf{I}_1 + \textbf{I}_1 \cdot \textbf{T}^{\rm(H_a)} \cdot \textbf{S} + a_F^{\rm(H_{m1},H_{m2})} \cdot \textbf{S} \cdot \textbf{I}_2 + \textbf{I}_2 \cdot \textbf{T}^{\rm(H_{m1},H_{m2})} \cdot \textbf{S}
\end{equation}
where $a_F^{\rm(H_a)}$ and \textbf{T}$^{\rm(H_a)}$ stand for the Fermi contact constant and the dipole-dipole interaction tensor for the acetylenic hydrogen nucleus, respectively, and $a_F^{\rm(H_{m1},H_{m2})}$ and \textbf{T}$^{\rm(H_{m1},H_{m2})}$ are averages of the coupling constants for the two methylenic hydrogen nuclei that are equivalent. In this manner, each energy level is denoted by six quantum numbers: $N$, $K_a$, $K_c$, $J$, $F_1$, and $F$.

\begin{table}
\small
\caption{Spectroscopic parameters of CH$_2$CCH (all in MHz).}
\label{table:constants}
\centering
\begin{tabular}{lcc}
\hline \hline
\multicolumn{1}{c}{Parameter}  & \multicolumn{1}{c}{Global Fit} & \multicolumn{1}{c}{\cite{Tanaka1997}} \\
\hline
$A$                          &  ~        288055\,$^a$       &  ~  288055           \\
$B$                          &  ~    9523.6746(41)    &  ~  9523.6775(60)     \\
$C$                          &  ~       9206.8776(41)       &  ~  9206.8805(60)     \\
$\Delta_N$                   &  ~        0.003004(72)       &  ~  0.003440(63)     \\
$\Delta_{NK}$                &  ~          0.3758(28)       &  ~  0.3753(28)       \\
$\Delta_K$                   &  ~        22.62\,$^a$        &  ~    22.62          \\
$\delta_N$                   &  ~        0.000103\,$^a$     &  ~    0.000103       \\
$\delta_K$                   &  ~        0.1575\,$^a$       &  ~    0.1575         \\
$\varepsilon_{aa}$           &  ~        $-$529.386(60)     &  ~  $-$529.386(60)   \\
$\varepsilon_{bb}$           &  ~         $-$11.524(31)     &  ~   $-$11.524(30)   \\
$\varepsilon_{cc}$           &  ~          $-$0.520(31)     &  ~    $-$0.520(30)   \\
$a_F$$^{\rm(H_a)}$           &  ~         $-$36.322(25)     &  ~   $-$36.323(24)   \\
$T_{aa}$$^{\rm(H_a)}$        &  ~          17.400(26)       &  ~    17.400(24)     \\
$T_{bb}$$^{\rm(H_a)}$        &  ~         $-$17.220(38)     &  ~   $-$17.220(37)   \\
$a_F$$^{\rm(H_m)}$           &  ~          $-$54.20(12)    &  ~    $-$54.21(11)   \\
$T_{aa}$$^{\rm(H_m)}$        &  ~         $-$14.122(20)    &  ~   $-$14.121(19)   \\
$T_{bb}$$^{\rm(H_m)}$        &  ~          12.88\,$^a$      &  ~    12.88          \\
$rms$\,$^b$                    &  ~               6.3         &  ~        7.0        \\
N\,$^c$                        &  ~               55          &  ~         46        \\
\hline
\end{tabular}
\tablenoteb{\\
Numbers in parentheses are 3$\sigma$ uncertainties in units of the last digits. H$\rm_a$ and H$\rm_m$ refer to the acetylenic and methylenic hydrogen nuclei, respectively.\\
$^a$\,Parameter fixed to the value reported by \cite{Tanaka1997}.\\
$^b$\,Standard deviation of the fit in kHz.\\
$^c$\,Number of lines included in the fit.\\
}
\end{table}

\begin{table}
\small
\caption{Rotational partition function ($Q_r$) of CH$_2$CCH at different temperatures.}
\label{table:qrot}
\centering
\begin{tabular}{cc}
\hline
\hline
Temperature (K)   & $Q_r$ \\
\hline
  9.375 &   294.5 \\
 18.750 &   718.9 \\
 37.500 &  1959.3 \\
 75.000 &  5506.5 \\
150.000 & 14710.5 \\
225.000 & 24377.0 \\
300.000 & 33479.9 \\
\hline
\end{tabular}
\end{table}

The results obtained from the fit are shown in Table~\ref{table:constants}, where they are compared with those reported by \cite{Tanaka1997}. As expected, the new derived parameters for CH$_2$CCH are almost identical to those reported before. The inclusion of the 5$_{0,5}$-4$_{0,4}$ transition in the fit only affects the rotational constants $B$ and $C$ and the distortion constants $\Delta_N$ and $\Delta_{NK}$. For $B$, $C$ and $\Delta_{NK}$ the differences are smaller than the 3$\sigma$ uncertainties. However, for the $\Delta_N$ distortion constant the difference is much larger, as expected, due to the inclusion of rotational transitions with higher quantum number $N$. We used the spectroscopic parameters obtained in this work for CH$_2$CCH to obtain accurate frequency predictions at mm wavelengths. The catalog file with the predicted frequencies and the calculated intensities at 300 K is provided at the CDS. The intensities are calculated adopting a dipole moment of 0.150 D, the experimental value measured by \cite{Kupper2002}. The rotational partition functions used in these predictions are listed at different temperatures in Table~\ref{table:qrot}. The rotational partition function was calculated considering a maximum value of 30 for the quantum number $N$.

\subsection{Excitation and abundance of CH$_2$CCH in \mbox{TMC-1}, and guidance for further searches}

\begin{figure}
\centering
\includegraphics[angle=0,width=0.95\columnwidth]{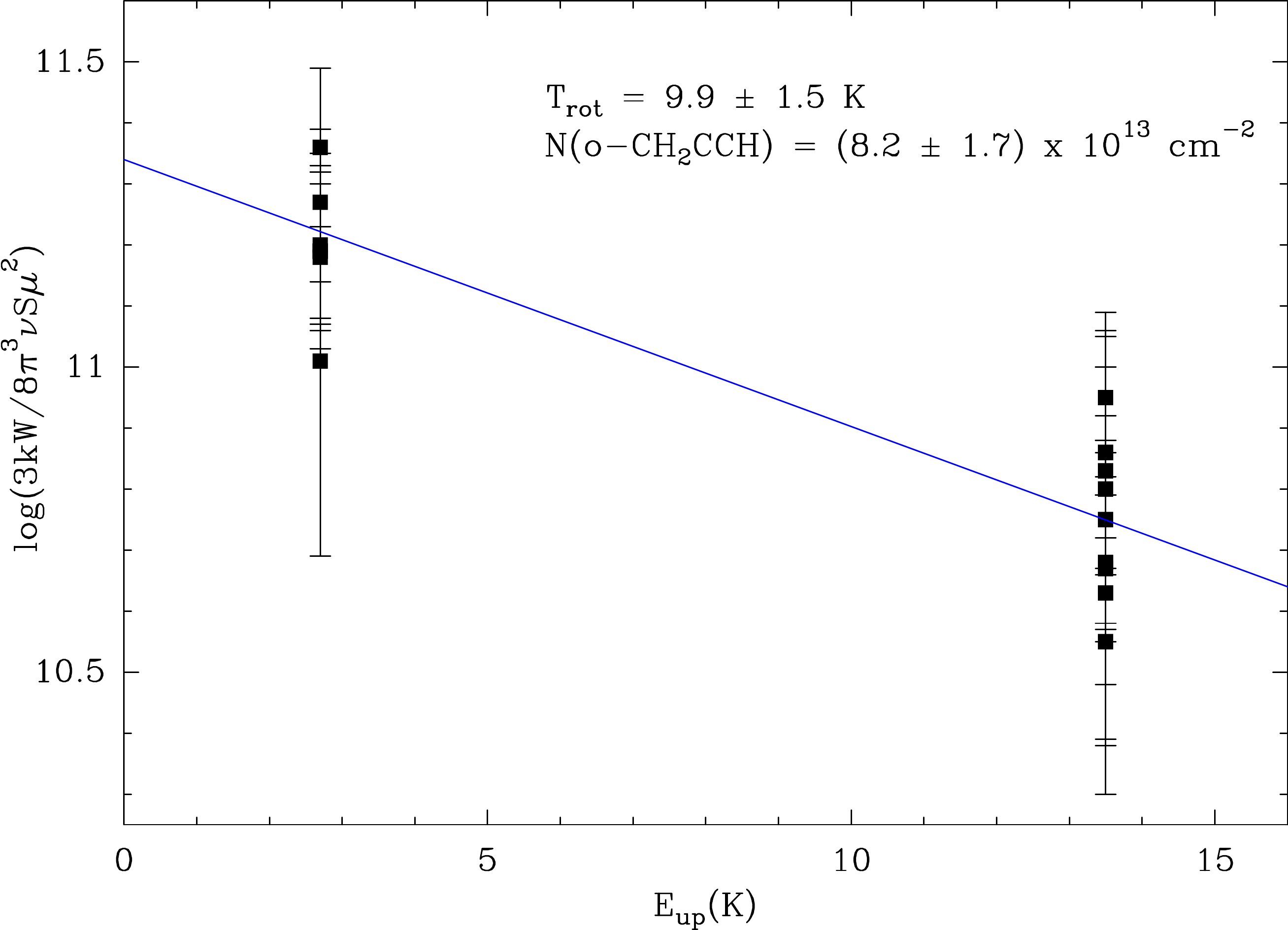}
\caption{Rotation diagram of ortho CH$_2$CCH in \mbox{TMC-1}.} \label{fig:rtd}
\end{figure}

We also present the observed spectrum of \mbox{TMC-1} at the frequency of the 2$_{0,2}$-1$_{0,1}$ transition of CH$_2$CCH (see bottom panel in Fig.~\ref{fig:lines}). This spectrum is more sensitive than that presented in \cite{Agundez2021a} because it includes additional observations taken with the Yebes 40m telescope. The rms noise level has decreased from 0.30 mK to 0.19 mK, per 38.15 kHz channel. As a consequence, the CH$_2$CCH lines are now more clearly detected. The new line parameters derived for the six hyperfine components of the 2$_{0,2}$-1$_{0,1}$ transition of CH$_2$CCH are given in Table~\ref{table:lines}.

As can be seen in Fig.~\ref{fig:lines}, the strongest hyperfine component of the 5$_{0,5}$-4$_{0,4}$ transition is about five times more intense than the strongest component of the 2$_{0,2}$-1$_{0,1}$. This is consistent with the rotational temperature of CH$_2$CCH being close to the gas kinetic temperature of \mbox{TMC-1}, $\sim$\,10~K \citep{Feher2016}. At this temperature, the 5$_{0,5}$ rotational level, with an energy of $\sim$\,13.5~K, is expected to be more populated than the 2$_{0,2}$ level, which has an energy of $\sim$\,2.7~K. In addition, the Einstein coefficient of spontaneous emission is about 20 times larger for the 5$_{0,5}$-4$_{0,4}$ transition than for the 2$_{0,2}$-1$_{0,1}$. These facts make the 5$_{0,5}$-4$_{0,4}$ transition at 93.6 GHz more favorable for detection than the 2$_{0,2}$-1$_{0,1}$ transition at 37.5 GHz. Indeed, if we assume typical values for the system temperatures, $T_{\rm sys}$ = 40 K at 37.5 GHz with the Yebes 40m telescope and $T_{\rm sys}$ = 120 K at 93.6 GHz with the IRAM 30m telescope, and we keep in mind that the line at 93.6 GHz is five times more intense than the 37.5 GHz line, the radiometer equation tells us that in order to detect the two lines with the same signal-to-noise ratio (S/N) one must invest $\sim$\,7 times more integration time at the Yebes 40m telescope than with IRAM 30m telescope. The fact that in our data the 2$_{0,2}$-1$_{0,1}$ transition is detected with similar or even higher S/N than the 5$_{0,5}$-4$_{0,4}$ transition (see Table~\ref{table:lines}) is a consequence of the much longer integration time invested with the Yebes 40m telescope (238 h) compared to that employed for the IRAM 30m spectrum (35.4 h). In summary, the rotational transitions in the $\lambda$\,3 mm wavelength band, in particular the 5$_{0,5}$-4$_{0,4}$ at 93.6 GHz (see below), are the most favorable for detection and should be the target in future searches for CH$_2$CCH in cold dark clouds.

The availability of two rotational transitions with different upper level energies allows us to constrain the rotational temperature of the propargyl radical in \mbox{TMC-1}. We built a rotation diagram using the velocity-integrated intensities given in Table~\ref{table:lines} and we derive a rotational temperature of 9.9\,$\pm$\,1.5~K (see Fig.~\ref{fig:rtd}). We therefore confirm the assumption made by \cite{Agundez2021a} that the rotational levels of CH$_2$CCH are thermalized at the gas kinetic temperature of \mbox{TMC-1}, $\sim$\,10~K \citep{Feher2016}. This fact is expected based on the low dipole moment of CH$_2$CCH (0.150 D; \citealt{Kupper2002}), which implies low critical densities, probably a few 10$^2$ cm$^{-3}$, i.e., well below the volume density of H$_2$ in \mbox{TMC-1}, a few 10$^4$ cm$^{-3}$ \citep{Pratap1997,Cordiner2013}. The column density derived from the rotation diagram for ortho CH$_2$CCH is (8.2\,$\pm$1.7)\,$\times$\,10$^{13}$ cm$^{-2}$. A more precise determination of the column density can be obtained by fitting the observed spectra with synthetic spectra calculated under local thermodynamic equilibrium. For this calculation we adopted a rotational temperature of 9.9\,K, as derived from the rotation diagram, a full width at half maximum (FWHM) of 0.72 km s$^{-1}$ for the 2$_{0,2}$-1$_{0,1}$ lines and 0.57 km s$^{-1}$ for the 5$_{0,5}$-4$_{0,4}$ lines, which are the arithmetic mean of the values derived for the hyperfine components of each transition (see Table~\ref{table:lines}), and assumed that the emission is distributed in the sky as a circle with a radius of 40\,$''$, as observed for various hydrocarbons in \mbox{TMC-1} \citep{Fosse2001}. The observed spectra at 37.5 GHz and 93.6 GHz are well reproduced adopting a column density of 7.5\,$\times$\,10$^{13}$ cm$^{-2}$ (see Fig.~\ref{fig:lines}). Assuming an ortho-to-para ratio of three, the column density of CH$_2$CCH (including ortho and para) in \mbox{TMC-1} is (1.0\,$\pm$0.2)\,$\times$10$^{14}$ cm$^{-2}$, which is slightly higher than the value derived previously by \cite{Agundez2021a}. The column density of the closed-shell counterpart CH$_3$CCH in \mbox{TMC-1} is (1.1-1.3)\,$\times 10^{14}$~cm$^{-2}$ \citep{Gratier2016,Cabezas2021}. Therefore, in this study we confirm that the propargyl radical is thermalized to the gas kinetic temperature of \mbox{TMC-1} and revise the abundance ratio CH$_2$CCH/CH$_3$CCH from slightly below one to nearly one.

There are other rotational transitions of CH$_2$CCH that lie in the frequency range covered by our Yebes 40m and IRAM 30m data. The two other transitions of ortho CH$_2$CCH that fall in the $\lambda$\,3 mm band, the 4$_{0,4}$-3$_{0,3}$ at 74.9 GHz and the 6$_{0,6}$-5$_{0,5}$ at 112.3 GHz, are predicted to be as intense as the 5$_{0,5}$-4$_{0,4}$. However, our data at these frequencies are not as sensitive as at 93.6 GHz, and thus the strongest hyperfine components of each transition are only marginally detected. System temperatures at 74.9 GHz and 112.3 GHz are higher than at 93.6 GHz, making the 5$_{0,5}$-4$_{0,4}$ transition the most favorable for detection. There are also several lines of para CH$_2$CCH accessible. Two of them, the 2$_{1,2}$-1$_{1,1}$ at 37.2 GHz and the 2$_{1,1}$-1$_{1,0}$ at 37.8 GHz, lie in the Q band and are covered by our Yebes 40m line survey, while two other transitions, the 5$_{1,5}$-4$_{1,4}$ at 92.8 GHz and the 5$_{1,4}$-4$_{1,3}$ at 94.4 GHz, lie in the $\lambda$\,3 mm band and are covered by our IRAM 30m telescope data. These lines are predicted to be less intense than those of ortho CH$_2$CCH, and thus are more difficult to detect. In our data only the strongest hyperfine components of the 5$_{1,5}$-4$_{1,4}$ and 5$_{1,4}$-4$_{1,3}$ transitions are barely visible. The S/N is however low and we have thus not attempted to fit them.

\section{Conclusions}

We detected the 5$_{0,5}$-4$_{0,4}$ transition of ortho CH$_2$CCH in \mbox{TMC-1} using the IRAM 30m telescope. The measured frequencies for ten hyperfine components of this transition are 0.2 MHz higher than the frequency predictions available in the CDMS catalog, a difference which is significant for radioastronomical purposes. We carried out a new spectroscopic analysis of the rotational spectrum of CH$_2$CCH in order to provide accurate frequencies at mm wavelengths. The intensity of the 5$_{0,5}$-4$_{0,4}$ transition, lying at 93.6 GHz, is $\sim$\,5 times higher in \mbox{TMC-1} than the 2$_{0,2}$-1$_{0,1}$ previously observed by \cite{Agundez2021a} using the Yebes 40m telescope. We conclude that a search for CH$_2$CCH in other cold interstellar sources should be carried out in the $\lambda$\,3 mm band, rather than at $\lambda$\,8 mm, where the telescope time investment is estimated to be about seven times cheaper. The rotational temperature of CH$_2$CCH in \mbox{TMC-1} is constrained to 9.9 $\pm$ 1.5 K, i.e., equal to the gas kinetic temperature, and the derived value of the column density is (1.0\,$\pm$\,0.2)\,$\times$\,10$^{14}$ cm$^{-2}$, which makes CH$_2$CCH one of the most abundant hydrocarbon radicals in \mbox{TMC-1}.

\begin{acknowledgements}

We acknowledge funding support from Spanish Ministerio de Ciencia e Innovaci\'on through grants PID2019-106110GB-I00, PID2019-107115GB-C21, and PID2019-106235GB-I00 and from the European Research Council (ERC Grant 610256: NANOCOSMOS). M.A. also acknowledges funding support from the Ram\'on y Cajal programme of Spanish Ministerio de Ciencia e Innovaci\'on (grant RyC-2014-16277). We thank the referee for a careful reading of the manuscript and for useful comments.

\end{acknowledgements}


\begin{thebibliography}{}

\bibitem[Ag\'undez et al.(2021a)]{Agundez2021a} Ag\'undez, M., Cabezas, C., Tercero, B., et al. 2021a, \aap, 647, L10 
\bibitem[Ag\'undez et al.(2021b)]{Agundez2021b} Ag\'undez, M., Marcelino, N., Tercero, B., et al. 2021b, \aap, 649, L4 
\bibitem[Botschwina et al.(1995)]{Botschwina1995} Botschwina, P., Oswald, R., Fl\"ugge, J., \& Horn, M. 1995, \zpc, 188, 29
\bibitem[Burkhardt et al.(2021)]{Burkhardt2021} Burkhardt, A. M., Lee, K. L. K., Changala, P. B., et al. 2021, \apj, 913, L18 
\bibitem[Cabezas et al.(2021)]{Cabezas2021} Cabezas, C., Endo, Y., Roueff, E., et al. 2021, \aap, 646, L1 
\bibitem[Cernicharo(1985)]{Cernicharo1985} Cernicharo, J. 1985, IRAM Internal Report 52
\bibitem[Cernicharo et al.(2012)]{Cernicharo2012} Cernicharo, J., Marcelino, N., Roueff, E., et al. 2012, \apjl, 759, L43
\bibitem[Cernicharo et al.(2020)]{Cernicharo2020} Cernicharo, J., Marcelino, N., Ag\'undez, M., et al. 2020, \aap, 642, L8 
\bibitem[Cernicharo et al.(2021a)]{Cernicharo2021a} Cernicharo, J., Ag\'undez, M., Cabezas, C., et al. 2021a, \aap, 647, L2 
\bibitem[Cernicharo et al.(2021b)]{Cernicharo2021b} Cernicharo, J., Cabezas, C., Ag\'undez, M., et al. 2021b, \aap, 647, L3 
\bibitem[Cernicharo et al.(2021c)]{Cernicharo2021c} Cernicharo, J., Ag\'undez, M., Cabezas, C., et al. 2021c, \aap, 649, L15 
\bibitem[Cernicharo et al.(2021d)]{Cernicharo2021d} Cernicharo, J., Ag\'undez, M., Kaiser, R. I., et al. 2021d, \aap, 652, L9 
\bibitem[Cernicharo et al.(2021e)]{Cernicharo2021e} Cernicharo, J., Ag\'undez, M., Kaiser, R. I., et al. 2021e, \aap, 655, L1 
\bibitem[Cordiner et al.(2013)]{Cordiner2013} Cordiner, M. A., Buckle, J. V., Wirstr\"om, E. S., et al. 2013, \apj, 770, 48
\bibitem[Feh\'er et al.(2016)]{Feher2016} Feh\'er, O., T\'oth, L. V., Ward-Thompson, D., et al. 2016, \aap, 590, A75
\bibitem[Foss\'e et al.(2001)]{Fosse2001} Foss\'e, D., Cernicharo, J., Gerin, M., \& Cox, P. 2001, \apj, 552, 168
\bibitem[Gratier et al.(2016)]{Gratier2016} Gratier, P., Majumdar, L., Ohishi, M., et al. 2016, \apjs, 225, 25
\bibitem[K\"upper et al.(2002)]{Kupper2002} K\"upper, J., Merritt, J. M., \& Miller, R. E. 2002, \jcp, 117, 647
\bibitem[Marcelino et al.(2007)]{Marcelino2007} Marcelino, N., Cernicharo, J., Ag\'undez, M., et al. 2007, \apj, 665, L127 
\bibitem[McGuire et al.(2018)]{McGuire2018} McGuire, B. A., Burkhardt, A. M., Kalenskii, S., et al. 2018, \science, 359, 202 
\bibitem[McGuire et al.(2020)]{McGuire2020} McGuire, B. A., Burkhardt, A. M., Loomis, R. A., et al. 2020, \apj, 900, L10 
\bibitem[McGuire et al.(2021)]{McGuire2021} McGuire, B. A., Loomis, R. A., Burkhardt, A. M., et al., 2021, \science, 371, 1265 
\bibitem[Miller \& Klippenstein(2001)]{Miller2001} Miller, J. A. \& Klippenstein, S. J. 2001, \jpca, 105, 7254
\bibitem[M\"uller et al.(2005)]{Muller2005} M\"uller, H. S. P., Schl\"oder, F., Stutzki, J., \& Winnewisser, G. 2005, \jmst, 742, 215
\bibitem[Pickett(1991)]{Pickett1991} Pickett, H. M. 1991, \jms, 148, 371
\bibitem[Pardo et al.(2001)]{Pardo2001} Pardo, J. R., Cernicharo, J., \& Serabyn, E. 2001, \ieeetap, 49, 1683
\bibitem[Pratap et al.(1997)]{Pratap1997} Pratap, P., Dickens, J. E., Snell, R. L., et al. 1997, \apj, 486, 862
\bibitem[Tanaka et al.(1997)]{Tanaka1997} Tanaka, K., Sumiyoshi, Y., Ohshima, Y., et al. 1997, \jcp, 107, 2728
\bibitem[Tercero et al.(2021)]{Tercero2021} Tercero, F., L\'opez-P\'erez, J. A., Gallego, et al. 2021, \aap, 645, A37
\bibitem[Zhao et al.(2021)]{Zhao2021} Zhao, L., Lu, W., Ahmed, M., et al. 2021, \sciadv, 7, eabf0360

\end{thebibliography}
\end{document}